\documentclass[aps,preprint,prd,showpacs,nofootinbib]{revtex4}

\usepackage{amsmath}
\usepackage{graphicx}
\usepackage{subfigure}
\usepackage{multirow}
\usepackage{bm}
\usepackage{amssymb}
\usepackage{mathtools}
\usepackage{enumerate}
\usepackage{color}
\usepackage[colorlinks,linkcolor=magenta,anchorcolor=cyan,citecolor=blue]{hyperref}
\usepackage{soul}

\begin{document}
\title{Alleviating both $H_0$ and $S_8$ tensions: early dark energy lifts the CMB-lockdown on ultralight axion}

\author{Gen Ye$^{1,2}$\footnote{yegen14@mails.ucas.ac.cn}}
\author{Jun Zhang$^{3,4}$\footnote{zhangjun@ucas.ac.cn}}
\author{Yun-Song Piao$^{1,2,3,5}$\footnote{yspiao@ucas.ac.cn}}

\affiliation{$^1$ School of Fundamental Physics and Mathematical
	Sciences, Hangzhou Institute for Advanced Study, UCAS, Hangzhou
	310024, China}

\affiliation{$^2$ School of Physics, University of Chinese Academy
	of Sciences, Beijing 100049, China}

\affiliation{$^3$ International Center for Theoretical Physics
	Asia-Pacific, Beijing/Hangzhou, China}

\affiliation{$^4$ Theoretical Physics, Blackett Laboratory,
	Imperial College, London, SW7 2AZ, UK}

\affiliation{$^5$ Institute of Theoretical Physics, Chinese
	Academy of Sciences, P.O. Box 2735, Beijing 100190, China}

\begin{abstract}

The existence of ultralight axion (ULA) with mass $\mathcal{O}(
10^{-26}\text{eV})$ is not favored by the CMB observations in the
standard $\Lambda$CDM model. We show that the inclusion of early
dark energy (EDE) will lift the CMB-lockdown on such ULA, and possibly other forms of dark matter beyond cold dark matter. By performing Monte Carlo Markov Chain
analysis, it is found that, as opposed to $\Lambda$CDM, the
AdS-EDE cosmology (with an Anti-de Sitter phase around
recombination) now allows the existence of axion with mass
$10^{-26}$ eV and predicts $6\%$ of the matter in our Universe to
be such ULA, which can also help alleviating the $S_8$ tension in EDE.

\end{abstract}

\maketitle

\section{Introduction}

Based on the standard cosmological constant cold dark matter ($\Lambda$CDM) model, the Planck cosmic
microwave background (CMB) data suggests that the current expansion
rate of the Universe (Hubble constant) is $H_0=67.4\pm0.5$km/s/Mpc
\cite{Planck:2018vyg}.
However, using Cepheid-calibrated supernovas, Riess et.al reported
$H_0=73.2\pm1.3$km/s/Mpc \cite{Riess:2020fzl}, in
$\sim4\sigma$ tension with the Planck result, see also other
independent local measurements, e.g.\cite{Birrer:2020tax}. It has
been widely thought that this so-called Hubble tension, if
confirmed\footnote{Recently, using Tip of the Red Giant Branch
(TRGB) calibrated supernova, Ref.\cite{Freedman:2021ahq} have
reported a lower value $H_0=69.8\pm0.6\text{(stat)}
\pm1.6\text{(sys)} $, statistically consistent with both CMB and
Cepheid values, which might imply uncounted uncertainties in
current local observations.}, likely signals new physics beyond
$\Lambda$CDM
\cite{Verde:2019ivm,Knox:2019rjx,DiValentino:2019qzk,Handley:2019tkm}.

The currently promising resolution of the Hubble tension is early
dark energy (EDE) \cite{Karwal:2016vyq,Poulin:2018cxd}. In
corresponding scenarios, the EDE component is non-negligible only
around matter-radiation equality before recombination, which
results in a suppressed sound horizon, so $H_0\gtrsim 70$km/s/Mpc.
In past years, the EDE models have been extensively studied
\cite{Agrawal:2019lmo,Lin:2019qug,Alexander:2019rsc,Smith:2019ihp,Niedermann:2019olb,Sakstein:2019fmf,Ye:2020btb,Ballesteros:2020sik,Braglia:2020bym,Chudaykin:2020acu,Ye:2020oix,Ye:2021nej,Chudaykin:2020igl,Lin:2020jcb,Sabla:2021nfy,Nojiri:2021dze,Karwal:2021vpk},
see also
Refs.\cite{Braglia:2020iik,Ballardini:2020iws,Braglia:2020auw} for early modified gravity, for thorough reviews see e.g
Refs.\cite{DiValentino:2021izs,Perivolaropoulos:2021jda}.
Furthermore, the existence of an Anti-de Sitter (AdS) phase around
recombination can allow a higher EDE fraction (without spoiling
fit to CMB, BAO and supernova light curves), so that the
corresponding AdS-EDE model can be $1\sigma$ consistent with local
$H_0$ measurements, see Refs.\cite{Ye:2020btb,Ye:2020oix}.

\begin{figure}[htp!]
    \centering
    \includegraphics[width=\linewidth]{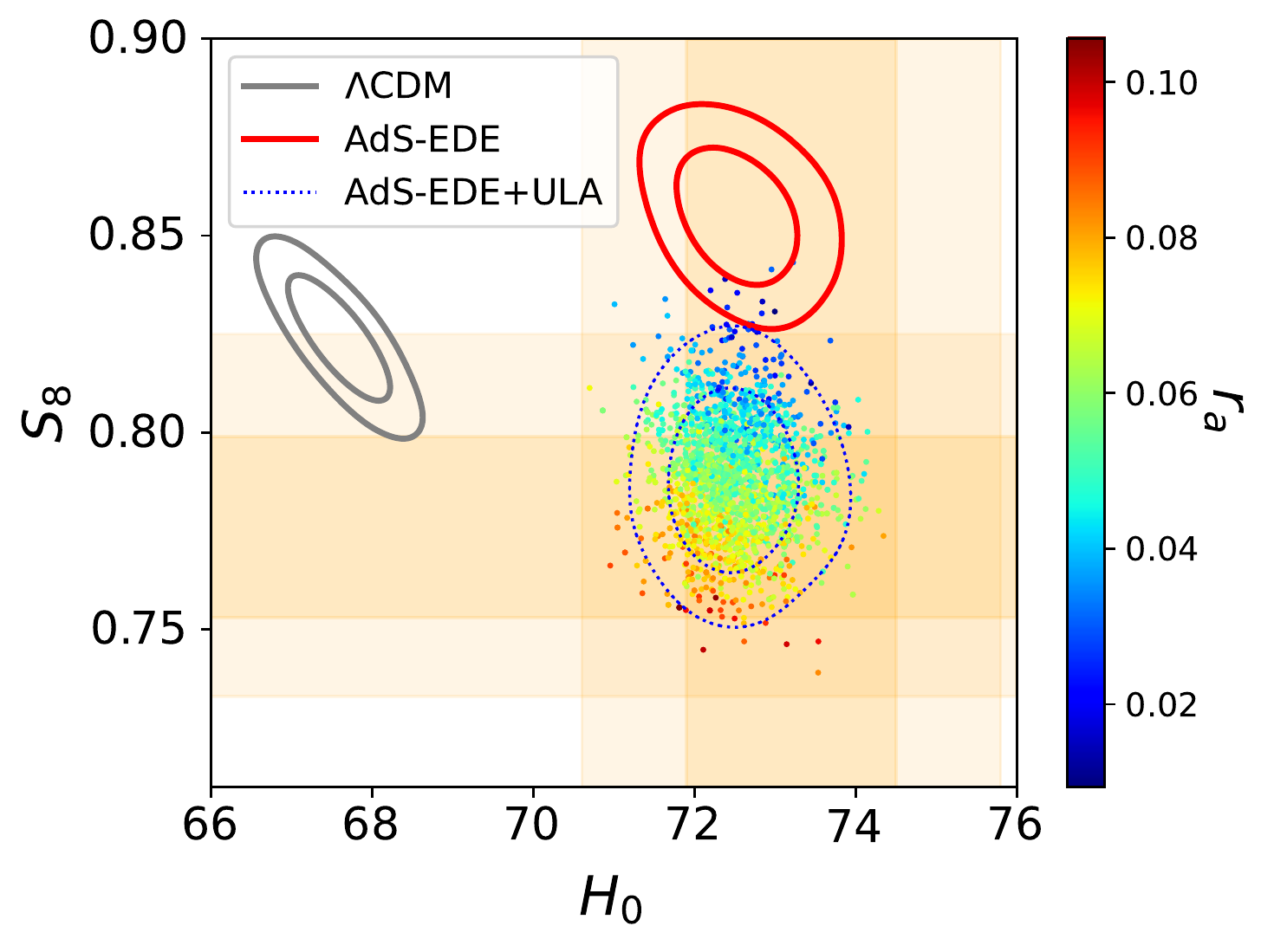}
\caption{Posterior distributions of $\Lambda$CDM, AdS-EDE and AdS-EDE with ultralight axion (AdS-EDE+ULA) cosmologies in the $H_0-S_8$ plain. Yellow bands show the $1\sigma$ and $2\sigma$ region of
local $S_8=0.755^{+0.019}_{-0.021}$ \cite{Asgari:2019fkq} and
$H_0=73.2\pm 1.3$ km/s/Mpc \cite{Riess:2020fzl} measurements.
Scattered points correspond to AdS-EDE+ULA with a color
coding for the energy fraction of ULA in total matter $r_a\equiv\Omega_{axion}/\Omega_{m}$.
}
    \label{H0-S8}
\end{figure}

In Ref.\cite{Ye:2020oix,Ye:2021nej}, it has been found that in
such pre-recombination solutions, the relevant cosmological
parameters must shift with $\delta H_0$, particularly the shift of
$\omega_m=\Omega_mh^2$ scales approximately as $\frac{\delta
\omega_{m}}{\omega_{m}}\sim 2\frac{\delta H_0}{H_0}$.  More
dust-like matter induces more clustering (perturbation growth) in
the matter-dominated era, so a higher $\sigma_8$ (the
amplitude of matter perturbations at $8h^{-1}$Mpc
scale) and $S_8\equiv\sigma_8\sqrt{\Omega_{m }/0.3}$. $S_8$ can be probed by local large scale structure (LSS) observations. Recently,
based on the $\Lambda$CDM model, the weak lensing measurements and
redshift surveys \cite{Asgari:2019fkq,KiDS:2020suj, Heymans:2020gsg, DES:2021wwk}
report lower $S_8$ values, see also \cite{Nunes:2021ipq}. At
first thought, it seems that the EDE resolutions of the Hubble tension
will inevitably suffer inconsistency with local
$S_8$ measurements
\cite{Hill:2020osr,Ivanov:2020ril,DAmico:2020ods,Jedamzik:2020zmd}
(except
\cite{Chudaykin:2020acu,Chudaykin:2020igl,Jiang:2021bab} for
combining Planck low-$\ell$ and SPTpol data), the so-called $S_8$
tension, because both larger $\omega_m$ and $n_s$ ($\delta
n_s\simeq 0.4\frac{\delta H_0}{H_0}$ \cite{Ye:2021nej}) will
enhance the matter power spectrum around $k_{8}\sim (8/h
\text{Mpc})^{-1}$, see also \cite{Vagnozzi:2021gjh}.

However, careful analysis of the physics behind the shift in $\omega_{m}$ tells a different story. The
$\omega_m$-$H_0$ correlation in Ref.\cite{Ye:2020oix} is actually
the consequence of $\theta_{CMB}$ and $\theta_{BAO}$ (angular
spacing of the acoustic peaks) after recombination being strictly constrained by observations, see also \cite{Pogosian:2020ded}, which,
however, only relies on the background evolution. In the standard CDM
model, $(\omega_m)_{bk}=(\omega_m)_{pt}$, where $(\omega_m)_{bk}$
is the fraction of pressureless ($p\simeq0$) energy
density (relevant to the background evolution) while $(\omega_m)_{pt}$ is the fraction of Jeans-unstable matter
(relevant to perturbation growth), thus
$\frac{\delta \omega_{m}}{\omega_{m}}\simeq 2\frac{\delta
H_0}{H_0}$ will suggest exacerbated $S_8$ tension for EDE. There
are, however, other possible DM forms which contribute to
$(\omega_m)_{bk}$ and $(\omega_m)_{pt}$ differently. We will show that EDE can indeed accommodate alternative form of DM, provided certain conditions are met, which will be clarified in the main text.

As an example, the alternative DM form we will consider is the ultra-light axion (ULA), which is theoretically well-motivated and naturally arise from
compactifications in string theory
\cite{Svrcek:2006yi,Arvanitaki:2009fg} and might constitute a
fraction of dark matter, see \cite{Hui:2021tkt} for a recent
review. It is interesting that axions with weak self-interaction
can lead to black hole superradiance, which can be constrained by
black hole spin measurements e.g.\cite{Davoudiasl:2019nlo},
polarimetric observations \cite{Chen:2019fsq}, and gravitational
wave observations \cite{Brito:2017wnc,Palomba:2019vxe,
Zhang:2018kib,Zhang:2019eid}. In addition, axion also could be
sourced by neutron stars
\cite{Sagunski:2017nzb,Huang:2018pbu,Zhang:2021mks}. It has been noticed in Ref.\cite{Allali:2021azp} that ULA of mass $10^{-26}$ eV might help alleviate the $S_8$ tension in EDE. However, Ref.\cite{Allali:2021azp} approximated EDE as an effective fluid with an instant transition in its equation of state parameter $w$ as well as the adiabatic sound speed $c_s^2=w$, which made it hard to see what exact role ULA plays in the cosmic evolution. Besides, the effect of EDE is most prominent when its energy fraction peaks immediately after thawing, at which time $w=\rm{const.}$ is not a good approximation since it is the transition phase from $w=-1$ to the final stage.

\begin{figure}
\includegraphics[width=0.8\linewidth]{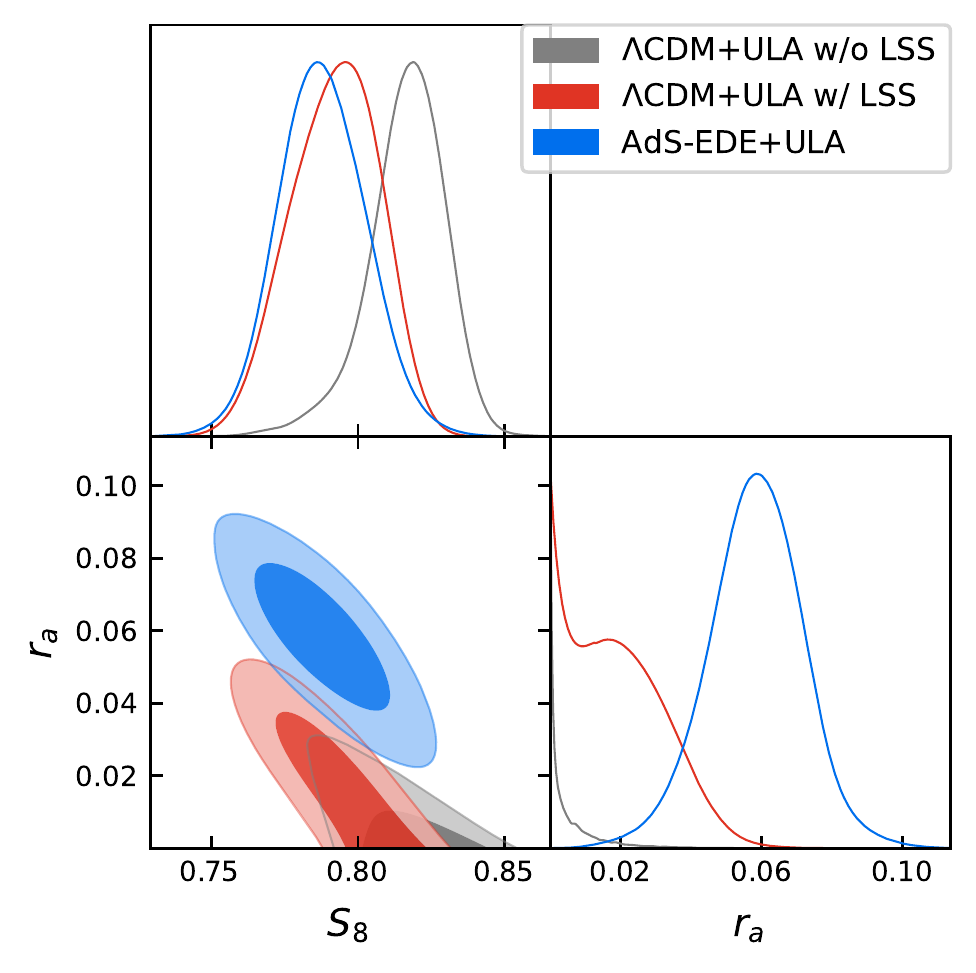}
\caption{68\% and 95\% posterior distribution of ULA energy
fraction $r_a\equiv\Omega_{axion}/\Omega_{m}$ and $S_8$ in
$\Lambda$CDM and AdS-EDE
using the baseline+fsBAO+$M_B$ dataset, see section-\ref{sec:model_data} for details. The $\Lambda$CDM+ULA w/o LSS result is obtained using P18+BAO+SN+$M_B$. Inclusion of $S_8$ related LSS information only biases the $\Lambda$CDM contour to a smaller
(larger) value of $S_8$ $(r_a)$ but does not change the existence
conclusion. The ULA used to produce these results has mass
$m_a=1.3\times10^{-26}$ eV and coupling constant
$f_a\sim10^{17}\text{GeV}$.} \label{raS8}
\end{figure}

In this letter we go one step further, treating both EDE and ULA robustly by solving the full background and linearly perturbed Klein-Gordon equations for both fields in the cosmology code\footnote{Our modified version of CLASS is available at \url{https://github.com/genye00/class_multiscf.git}}. With this improved method, we identified the core DM properties required by EDE to maintain good fit to CMB, which explains why ULA with mass $10^{-26}$ eV can help with the $S_8$ tension in EDE. More importantly, our analysis highlights the essential role of EDE in opening the parameter space for alternative DM forms, particularly ULA. The existence of
ULA with mass $10^{-26}$ eV is in fact not favored by $\Lambda$CDM, even if the full $S_8$ related information is included, see
\cite{Hlozek:2017zzf,Lague:2021frh} and also Fig.\ref{raS8}. As we will show in the main text, it is EDE that lifts the CMB lockdown on such DM form.

\section{Model and datasets}\label{sec:model_data}

As a phenomenological model, the AdS-EDE potential we will be working
with is $V(\phi)=V_0(\phi/M_p)^{4}-V_{ads}$, which is glued to a
cosmological constant $V(\phi)=\Lambda$ by interpolation
\cite{Ye:2020btb}, where $V_{ads}>0$ is the AdS depth. It is
parameterized by $z_c$ (the redshift at which the field starts
rolling), $f_{ede}(z_c)$ (the energy fraction of EDE at $z_c$) and
fixed\footnote{According to Ref.\cite{Ye:2020oix}, the existence of an AdS phase is weakly hinted, see also Appendix-\ref{apdx:mcmc_releasepar}, thus to explore the effect of AdS-EDE we choose to fix the AdS depth $\alpha_{ads}$ to the value in Ref.\cite{Ye:2020btb}, which is compatible with P18+BAO+SN.} $\alpha_{ads}\equiv V_{ads}/H(z_c)=3.79\times10^{-4}$. Non-zero $\alpha_{ads}$ value effectively puts a theoretical lower bound on $f_{ede}$ and cuts out part of the viable parameter space due to the fact that the field would be trapped in the disastrous negative energy region if $f_{ede}$ becomes too small. The axion potential is
$V(a)=m_a^2f_a^2[1-\cos(a/f_a)]$ with the coupling constant $f_a<M_p$.
Theoretically, the initial phase $\Theta_i\equiv a_i/f_a$ of the
axion is arbitrary. To reduce the number of free parameters, we fix\footnote{ULA thaws relatively deep ($z\sim17000$ for the case we studied) in the radiation dominant era, thus $\Theta_i$ should only have subdominant effect on the observed CMB provided $\Theta_i$ does not take extreme values close to $\pi$ or $0$. See Appendix-\ref{apdx:mcmc_releasepar} for more details.} the initial phase to the general value $\Theta_i=2$.

Axion-like scalar field rapidly oscillating around the minimum of
its potential mimics a $w=0$ fluid at the background level while
on the perturbation level, it can be effectively regarded as
perfect fluid with $c_s^2\sim\frac{k^2}{4m_a^2a^2}$, corresponding
to the Jeans scale $k_J/a\sim6^{1/4}\sqrt{Hm_a}$
\cite{Hu:2000ke,Hwang:2009js,Marsh:2010wq}. A part of
$(\omega_m)_{bk}$ might be constituted of axion
$\omega_a=r_a(\omega_m)_{bk}$. To have axion contribute
differently to $(\omega_m)_{bk}$ and $(\omega_m)_{pt}$ at the $S_8$
scale, we must at least have $k_J\lesssim k_8$ at matter-radiation
equality, which thus suggests the axion mass
$m_a\lesssim\mathcal{O}(10^6H_0)\sim\mathcal{O}(10^{-26}\text{eV})$.
The lower bound on $m_a$ is set by requiring that the axion field
mimics matter ($w=0$) at low redshift. To have an order of
magnitude estimation, we require $m_a>100H(z=10)$ or equivalently
$m_a\gtrsim2\times10^3H_0\sim4h\times10^{-30}\text{eV}$.

\begin{figure}
    \centering
    \includegraphics[width=0.8\linewidth]{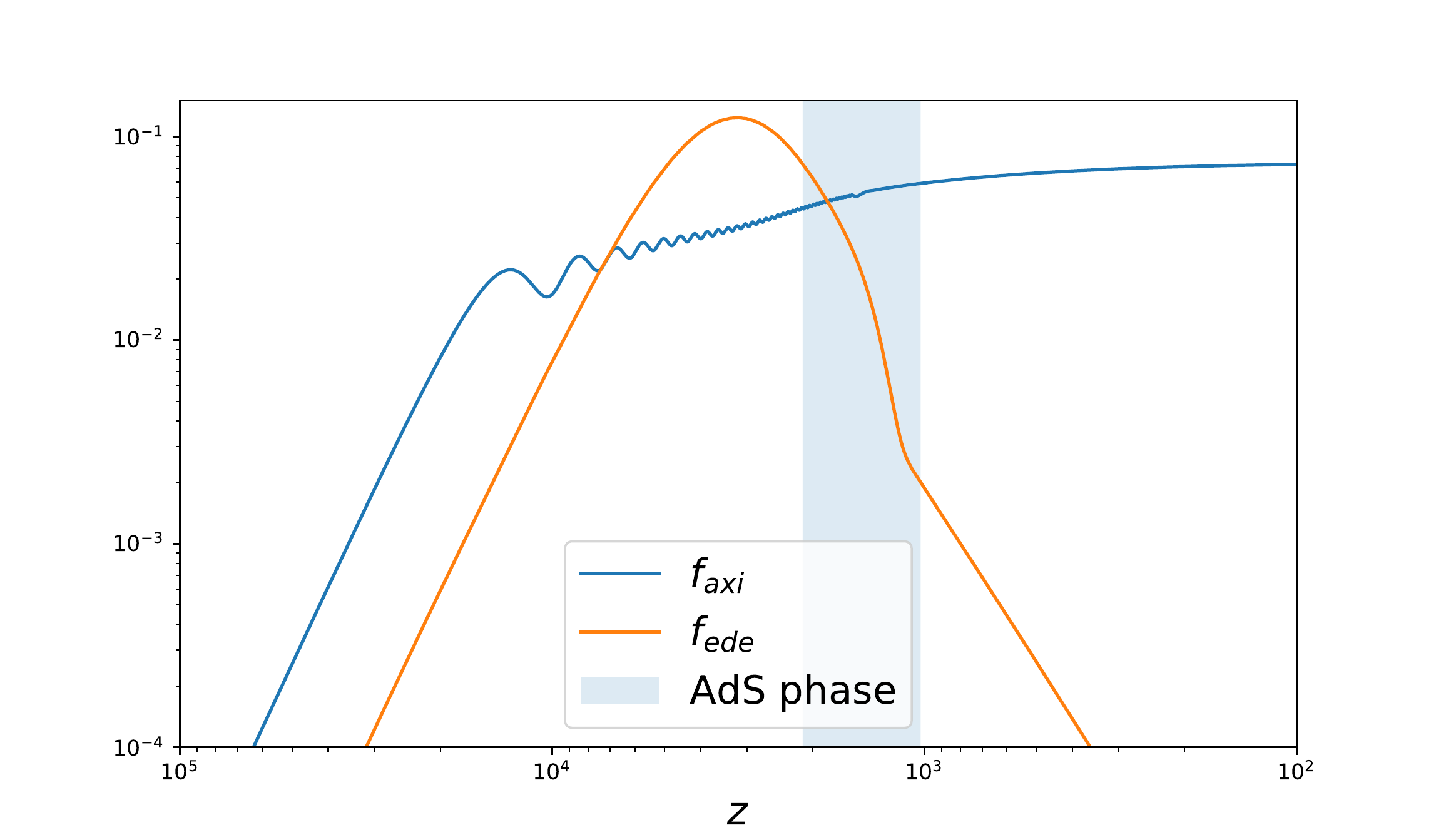}
\caption{Evolution of the energy
fractions $f_{ede}$ and $f_{axi}$ with
respect to redshift in the bestfit AdS-EDE+ULA model. Axion and EDE fields become dynamical at $z\simeq1.7\times10^4$ and $z\simeq4000$ respectively.}
    \label{fefa}
\end{figure}

In Fig.\ref{fefa}, we plot the evolution of $f_{ede}(z)$ and
$f_{axi}(z)$ (the energy fraction of axion), respectively. Here, the EDE
field responsible for resolving the Hubble tension thaws at $z=a \
few \ thousand$, while the axion (equivalently another EDE field with thawing time determined by $m_a=3H(z)$)
making $(\omega_m)_{bk}\neq (\omega_m)_{pt}$ at the $S_8$ scale thaws
well before matter-radiation equality, see also recent
Refs.\cite{Fung:2021wbz,Fung:2021fcj} for axion thawing after
recombination with $m_a\sim10^{-29}\text{eV}$.

To calculate linear cosmological perturbations and perform the
Monte Carlo Markov Chain (MCMC) analysis, we modified
Montepython-3.4 \cite{Audren:2012wb,Brinckmann:2018cvx} and CLASS
\cite{Lesgourgues:2011re,Blas:2011rf} to implement the AdS-EDE
model and multiple axion species. In particular, for ULA,
we solve the full Klein-Gordon equation until it enters the
rapidly oscillation phase, specified by $m_a/H\gg1$, after that we
follow
Refs.\cite{Hu:2000ke,Hwang:2009js,Marsh:2010wq,Poulin:2018dzj} and
adopt the axion field fluid approximation. Our datasets include:
\begin{enumerate}
\item \textbf{baseline:} \textbf{P18+SN+BAOLz+$\bm{S_8}$}.

\textbf{P18}: The most recent Planck2018 high-$l$ TTTEEE
likelihood together with low-$l$ TT, EE and Planck lensing data
\cite{Planck:2019nip}. \\
\textbf{SN}: The Pantheon supernova data \cite{Scolnic:2017caz}.
\\
\textbf{BAOLz}: The post-reconstructed small-$z$ BAO data from 6dF
\cite{Beutler:2011hx} and MGS \cite{Ross:2014qpa}.\\
$\mathbf{S_8}$: The KiDS+VIKING+DESY1 \cite{Asgari:2019fkq}
combined result $S_8=0.755^{+0.019}_{-0.021}$ is regarded as a
Gaussian prior\footnote{The more recent KiDS-1000 result gives $S_8=0.759^{+0.024}_{-0.021}$ \cite{KiDS:2020suj} and $S_8=0.766^{+0.02}_{-0.014}$ for KiDS-1000+BOSS+2dFLenS \cite{Heymans:2020gsg}. The new DES-Y3 reports $S_8=0.776\pm0.017$ \cite{DES:2021wwk}. The Gaussian prior we use is statistically compatible with these updated results.}.

\item \textbf{baseline+EFT/fsBAO+}$\bm{M_B}$.

\textbf{EFT}: The full shape galaxy power spectrum data from BOSS
DR12 extracted by EFTofLSS \cite{Baumann:2010tm,Carrasco:2012cv}
(with the publicly available pyBird code
\cite{DAmico:2020kxu,Colas:2019ret}), as well as the
post-reconstructed BAO data with corresponding covariance matrix.\\
\textbf{fsBAO}: BOSS DR12 post-reconstructed high-$z$ BAO and
redshift space distortion $f\sigma_8$ data with its covariance
matrix \cite{BOSS:2016wmc}, not independent of EFT. \\
$\bm{M_B}$: In light of
Refs.\cite{Lemos:2018smw,Benevento:2020fev,Camarena:2021jlr,Efstathiou:2021ocp,Freedman:2021ahq},
the SH0ES result \cite{Riess:2020fzl}, the absolute magnitude of
supernovas $M_B=-19.224\pm0.042$ mag, is regarded as a Gaussian
prior\footnote{In light of Ref.\cite{Efstathiou:2021ocp}-v1 in
arXiv, we take $M_B=-19.224\pm0.042$ mag to perform all the MCMC
analysis, but in v3 it is changed to $M_B=-19.214\pm0.037$ mag. Also, during the revision process of this paper, the SH0ES group reported new result $M_B=-19.253\pm0.027$ \cite{Riess:2021jrx} with Pantheon+\cite{Scolnic:2021amr}. The difference is statistically marginal and should not
significantly affect our results.}.
\end{enumerate}

\section{Results and analysis} \label{sec:result}

We fix\footnote{This accelerates convergence and is sufficient for our purpose of studying the cosmological effect of $\sim10^{-26}$ eV ULA. See Apendix-\ref{apdx:mcmc_releasepar} for results of varying $m_a$.} the ULA mass to its bestfit value $m_a=8.6\times10^6H_0\simeq1.8h\times10^{-26}\text{eV}$ in AdS-EDE+ULA cosmology with the baseline+EFT+$M_B$ datasets and
perform MCMC analysis over the cosmological parameter set
$\{\omega_b,\omega_{cdm},H_0,\tau_{rion},\ln10^{10}A_s,n_s,f_{ede}(z_c),\ln(1+z_c),f_a\}$.

\begin{table}
    \begin{tabular}{|c|c|c|c|c|}
    \hline
    \multirow{2}{*}{Parameters}&$\Lambda$CDM&AdS-EDE&\multicolumn{2}{c|}{AdS-EDE+ULA}\\
    \cline{2-5}
    &\multicolumn{3}{c|}{baseline+EFT+$M_B$}&baseline+fsBAO+$M_B$\\ \hline
    $100~\omega{}_{b }$&$2.248(2.243)_{-0.015}^{+0.014}$& $2.348(2.347)_{-0.015}^{+0.016}$&$2.338(2.331)_{-0.016}^{+0.018}$ &$2.335(2.358)_{-0.017}^{+0.018}$\\
    $\omega{}_{cdm }$ &$0.1178(0.118)_{-0.00082}^{+0.00078}$&$0.1312(0.1318)_{-0.0017}^{+0.0018}$& $0.1242(0.1229)_{-0.0025}^{+0.0023}$& $0.1231(0.1239)_{-0.0025}^{+0.0024}$\\
    $H_0$&$68.81(68.66)_{-0.38}^{+0.4}$ &$73.19(73.32)_{-0.53}^{+0.51}$&$72.53(72.14)_{-0.58}^{+0.51}$ & $72.48(73.01)_{-0.57}^{+0.52}$ \\
    $\ln10^{10}A_{s }$ &$3.034(3.038)_{-0.015}^{+0.015}$ &$3.054(3.051)_{-0.014}^{+0.016}$&$3.063(3.066)_{-0.015}^{+0.014}$& $3.065(3.071)_{-0.014}^{+0.014}$ \\
    $n_{s }$& $0.9688(0.9668)_{-0.0038}^{+0.0037}$ &$0.9976(0.9976)_{-0.0039}^{+0.0042}$&$0.9885(0.9856)_{-0.0048}^{+0.0048}$ & $0.9885(0.991)_{-0.0048}^{+0.005}$ \\
    $\tau{}_{reio }$ & $0.0533(0.0564)_{-0.008}^{+0.007}$ &$0.0499(0.0488)_{-0.0075}^{+0.0079}$& $0.0534(0.0549)_{-0.0076}^{+0.007}$ &$0.0540(0.0584)_{-0.0074}^{+0.0072}$ \\
    \hline
    $\ln(1+z_{c})$&- &$8.258(8.233)_{-0.097}^{+0.1}$&$8.255(8.256)_{-0.069}^{+0.088}$& $8.246(8.293)_{-0.068}^{+0.082}$ \\
    $f_{ede }$&- &$0.1068(0.1083)_{-0.0063}^{+0.0043}$&$0.1094(0.1124)_{-0.0064}^{+0.0036}$ & $0.1116(0.1203)_{-0.0084}^{+0.0036}$  \\
    $\alpha_{ads}$&-&$3.79\times10^{-4}$&$3.79\times10^{-4}$&$3.79\times10^{-4}$\\
    \hline
    $m_a \ (10^6H_0)$ &-&-&$8.6$&$8.6$\\
    $f_{a }/M_p$ &-& -&$0.0541(0.0631)_{-0.0057}^{+0.0081}$& $0.0552(0.0597)_{-0.0055}^{+0.0077}$ \\
    $r_a$&-&-&$0.0567(0.0751)_{-0.013}^{+0.015}$&$0.0585(0.0674)_{-0.013}^{+0.015}$\\
    \hline
    $M_{B }$&$-19.39(-19.39)_{-0.011}^{+0.011}$ &$-19.26(-19.25)_{-0.015}^{+0.015}$&$-19.27(-19.28)_{-0.016}^{+0.014}$ & $-19.27(-19.26)_{-0.016}^{+0.014}$ \\
    $S_8$& $0.808(0.811)_{-0.009}^{+0.008}$ &$0.836(0.837)_{-0.01}^{+0.01}$&$0.785(0.772)_{-0.016}^{+0.015}$ &$0.788(0.776)_{-0.016}^{+0.015}$ \\
    $\Omega_{m }$ & $0.2964(0.2978)_{-0.0049}^{+0.0048}$&$0.2888(0.2888)_{-0.0049}^{+0.0045}$& $0.2975(0.3037)_{-0.0057}^{+0.0053}$ & $0.2993(0.2966)_{-0.0055}^{+0.0052}$ \\
    \hline
    \end{tabular}
\caption{Mean and $1\sigma$ values of cosmological parameters in AdS-EDE+ULA, as well as the original AdS-EDE
model in Ref.\cite{Ye:2020btb} and the standard $\Lambda$CDM
model. The axion mass is $m_a\simeq1.8h\times10^{-26}$eV. Bestfit
values are reported in the parenthesis.} \label{early_par}
\end{table}

\begin{table}
    \begin{tabular}{|c|c|c|c||c|}
    \hline
    \multirow{2}{*}{Dataset}&$\Lambda$CDM&AdS-EDE&\multicolumn{2}{c|}{AdS-EDE+ULA}\\
    \cline{2-5}
    &\multicolumn{3}{c||}{baseline+EFT+$M_B$}&baseline+fsBAO+$M_B$\\ \hline
    Planck18 high-$l$ TTTEEE&2347.70&2354.54 &2352.72 &2353.82 \\
    \hline
    Planck18 low-$l$ TT&22.87&20.39 &20.71 &20.87 \\
    \hline
    Planck18 low-$l$ EE&396.32&395.98 &396.06 &396.59 \\
    \hline
    Planck18 lensing&10.31&10.54 &9.82 &9.75 \\
    \hline
    Pantheon&1027.12&1027.04 & 1026.94&1026.89 \\
    \hline
    BAO low $z$&2.24&3.10 &1.53 &2.11 \\
    \hline
    BAO high $z$+ $f\sigma_8$&-&-& -&7.14 \\
    \hline
    eft withbao highzNGC&66.48&65.84&69.72&-\\
    \hline
    eft withbao highzSGC&62.92&64.51&60.73&-\\
    \hline
    eft withbao lowzNGC&71.14&71.39&71.26&-\\
    \hline
    $S_8$&7.89&16.61 &1.18&1.06 \\
    \hline
    $M_B$&12.00&0.05 & 0.79&0.12 \\
    \hline
    \hline
    $\chi^2_{CMB}$&2777.2&2781.45&2779.31&2781.03\\
    $\chi^2_{EFT}$&200.54&201.74&201.71&-\\
    $\chi^2_{tot}$&4026.99&4029.99&4011.46&3818.35\\
    \hline
    \end{tabular}
\caption{Bestfit $\chi^2$ per dataset for AdS-EDE, AdS-EDE+ULA and $\Lambda$CDM. The axion mass is $m_a\simeq1.8h\times10^{-26}$eV.}
\label{early_chi2}
\end{table}

We present in Table-\ref{early_par} the mean and bestfit values of
cosmological parameters in the best constrained models. For AdS-EDE+ULA, we also present results for baseline+fsBAO+$M_B$, which gives very similar cosmological constraints to baseline+EFT+$M_B$ but converges significantly faster than the latter due to no need of computing loop corrected matter power spectrum for the EFT likelihood. The corresponding
bestfit $\chi^2$ per dataset is reported in
Table-\ref{early_chi2}. The bestfit model implies an axion field thawing at $z\simeq1.7\times10^4$ with mass
$m_a\simeq1.8h\times10^{-26}$eV and coupling constant
$f_a\simeq1.5\times10^{17}$GeV, corresponding to the
axion-to-matter ratio $r_a\equiv\Omega_{axion}/\Omega_m\simeq6.7\%$.

\begin{figure}
    \centering
    \includegraphics[width=0.8\linewidth]{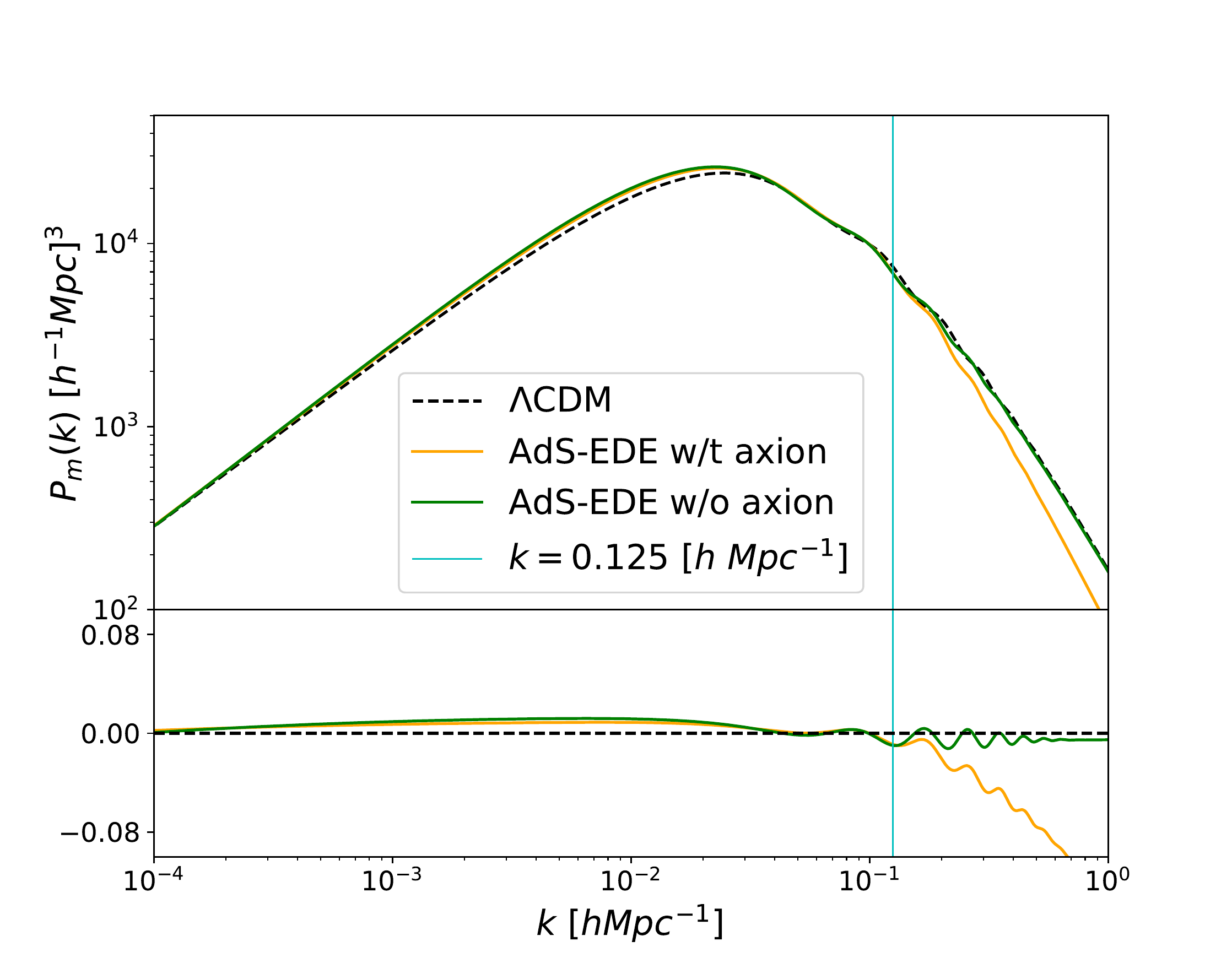}
\caption{In the bestfit AdS-EDE and AdS-EDE+ULA models,
linear matter power spectrum at $z=0$ is compared with that in the
$\Lambda$CDM model.}
    \label{early_pk}
\end{figure}

In Fig.\ref{H0-S8}, we see that the addition of an axion field
effectively lowers the predicted $S_8$ value, without interfering
with the ability of AdS-EDE to solve the Hubble tension.
Interestingly, now $H_0=72.53_{-0.58}^{+0.51}$ and
$S_8=0.785_{-0.016}^{+0.015}$, both are within $1\sigma$ bands of the local
measurements. According to the baseline+EFT+$M_B$ columns in Table.\ref{early_chi2}, as opposed to previous studies using P18+BAO+SN \cite{Ye:2020btb,Ye:2020oix}, AdS-EDE now yields obviously worse fit to the CMB data ($\Delta\chi^2_{CMB}=+4.25$) as well as $S_8$ prior ($\Delta \chi_{S_8}^2=+8.72$) compared with $\Lambda$CDM, due to the inclusion of LSS related datasets (EFT and $S_8$) and the exacerbated $S_8$ tension generic to EDE-like models \cite{Ivanov:2020ril,DAmico:2020ods}, see also Fig.\ref{H0-S8}. Compared with AdS-EDE alone, inclusion of ULA alleviates the tension in data due to improved $\chi^2_{S_8}$ and $\chi^2_{CMB}$. Despite the improvement over AdS-EDE alone, AdS-EDE+ULA still does not fit the CMB data as well as $\Lambda$CDM, implying some residual tension. To quantify the residual tension, we use the metric $Q^2_{DMAP}\equiv \chi^2_{\rm{min}, D_1+D_2}-\chi^2_{\rm{min}, D_1}-\chi^2_{\rm{min}, D_2}$ from \cite{Raveri:2018wln, Schoneberg:2021qvd}, where $D_1$ and $D_2$ are two datasets. Taking $D_1$=P18+BAOLz+SN+EFT and adding the $S_8$ and $M_B$ prior one at a time we get $Q_{DMAP}=4.5$ for $D_2=S_8$, $Q_{DMAP}=0.9$ for $D_2=M_B$ and $Q_{DMAP}=5.1$ for $D_2=S_8+M_B$ in the AdS-EDE+ULA model, where we have assumed $\chi^2_{\rm{min}, D_2}=0$ since $D_2$ contains at most two data points. See Appendix-\ref{apdx:mcmc_s8mb} for details. $r_a$ is clearly negatively
correlated with $S_8$, since larger $r_a$ will bring larger
suppression of matter perturbations on the $S_8$ scale, see
Fig.\ref{early_pk}.

\begin{figure}
    \centering
    \includegraphics[width=\linewidth]{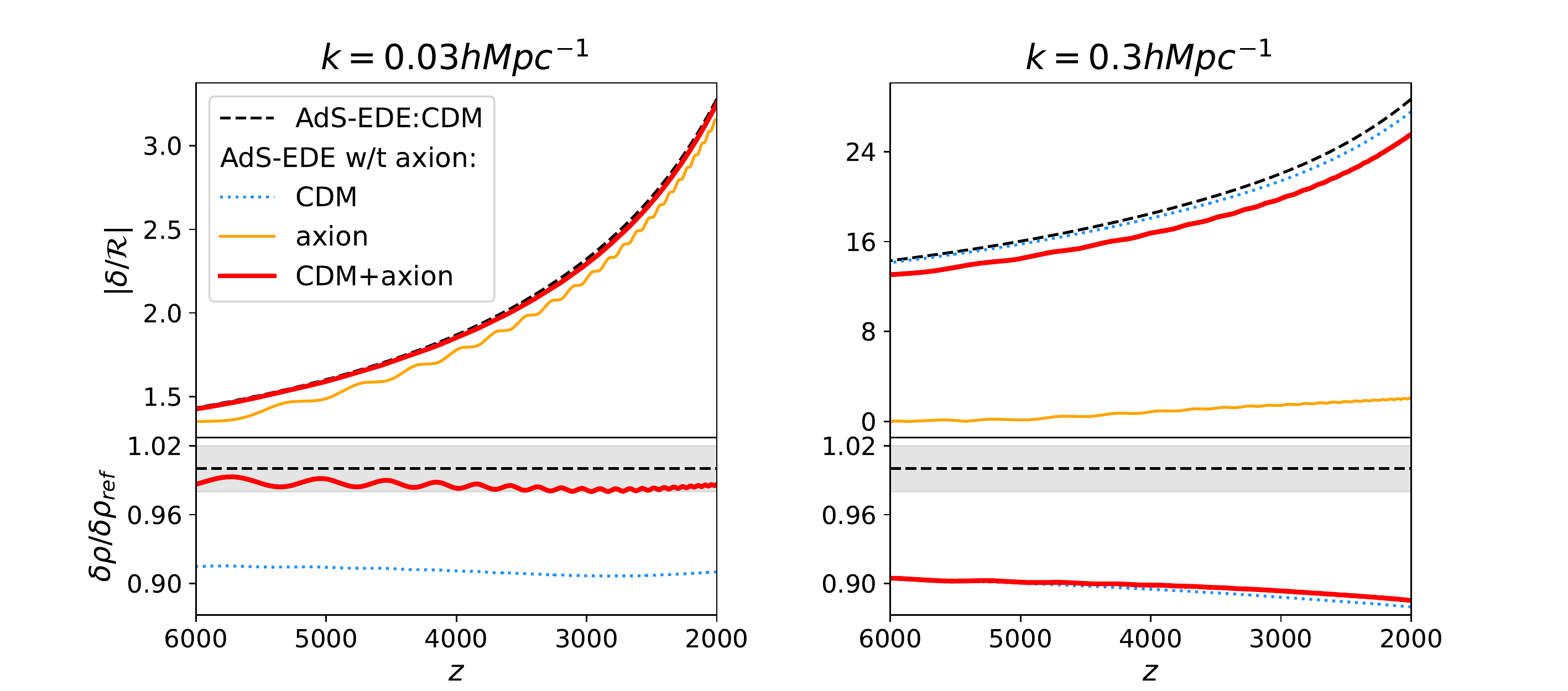}
\caption{Density perturbation for axion and CDM in the bestfit
AdS-EDE and AdS-EDE+ULA models. Upper panels plot the
relative density perturbation $\delta\equiv\delta\rho/\rho$,
normalized to the primordial curvature perturbation $\mathcal{R}$.
Lower panels plot the density perturbations with $\delta\rho_{ref}\equiv\delta\rho_{cdm}$ in the AdS-EDE
model. Gray band stands for $2\%$ relative variation around the
bestfit AdS-EDE model.}
\label{delta_ca}
\end{figure}

In Table-\ref{early_par}, the AdS-EDE+ULA model predicts lower $\omega_{cdm}\simeq0.124$ compared with
$\omega_{cdm}>0.13$ in the original AdS-EDE. As
pointed out by Ref.\cite{Ye:2020oix}, the background CMB+BAO compatibility requires a
larger $\omega_m$. This must be achieved in a way compatible with
the full anisotropic CMB dataset. EDE adds to the total amount of Jeans
stable species before recombination. Thus to have the correct amount of
radiation driving and early ISW, the amount of clustering species has to be increased accordingly
\cite{Ye:2020oix,Vagnozzi:2021gjh}. Assuming DM is entirely consisted of CDM, the majority of the needed increase in clustering species then must be realized by CDM, shifting $\omega_{cdm}$ in EDE cosmologies to a considerably larger vale than $\Lambda$CDM, enhancing the matter power spectrum on small scales and exacerbating the $S_8$ tension. This is identified as a difficulty of EDE in Ref.\cite{Hill:2020osr,Ivanov:2020ril,DAmico:2020ods,Jedamzik:2020zmd}. However, from another point of view, this might also be an implication of additional DM forms beyond CDM in EDE. The key property of these potential new DM forms is that they must act as Jeans unstable species on the relevant scales to compensate for the EDE pressure support. In AdS-EDE+ULA, this role is played by the ULA, thus a
lower $\omega_{cdm}$ is predicted. Additionally, ULA behaves as matter on the background level (i.e. $\rho_a\propto a^{-3}$) so $\Omega_{m}$ is nearly unchanged compared with $\Lambda$CDM in Table.\ref{early_par}, saturating the bound reported in Ref.\cite{Ye:2020oix}. The effect of EDE is
only significant at scales entering horizon near $z_c$ when the
EDE field is non-negligible, roughly corresponding to the first
acoustic peak ($l\lesssim300$). Thus one expects $
k_{J,eq}\gtrsim\mathcal{O}(0.01)\text{Mpc}^{-1}$ which disfavors
$m_a\lesssim\mathcal{O}(10^{-27}\text{eV})$. Recall from previous section that for ULA to have effect on the $S_8$ tension, we need $m_a\lesssim \mathcal{O}(10^{-26}\rm{eV})$, implying a mass range of interest $\mathcal{O}(10^{-26}\rm{eV})\gtrsim m_a\gtrsim\mathcal{O}(10^{-27}\rm{eV})$. The axion mass
$m_a=8.6\times10^6H_0\simeq1.8h\times10^{-26}\text{eV}$ used to
perform the MCMC analysis is within this mass range and corresponds to $k_{J,eq}\sim
0.16h\text{Mpc}^{-1}$, well above the EDE relevant scale. In
Fig.\ref{delta_ca}, we clarify this keypoint by plotting the
density perturbations for axion and CDM near $z_{eq}$ for
$k_{EDE}=0.03h\text{Mpc}^{-1}$ and $k_{LSS}=0.3h\text{Mpc}^{-1}$,
respectively. At $k_{EDE}$, axion mimics CDM in terms of perturbation growth and
together with the actual CDM, they produce correct amount of
$\delta\rho$ for clustering species as required by the AdS-EDE
bestfit (left panel of Fig.\ref{delta_ca}). Around the LSS
relevant scales, axion suppresses clustering (right
panel of Fig.\ref{delta_ca}).
%

\section{Conclusion} \label{sec:conclusion}

We argued that the increase in $\omega_{m}$ and clustering species required by EDE might be a possible indication of additional DM forms beyond CDM. We studied a well-established example of DM, the ULA, in EDE and showed explicitly by performing MCMC analysis that, as opposed to $\Lambda$CDM, AdS-EDE can indeed accommodate non-negligible amount ($\sim6\%$) of ULA with mass $m_a\simeq1.3\times10^{-26}$ eV and coupling constant
$f_a\simeq1.5\times10^{17}$ GeV. Interestingly, existence of such ULA species is not favored by the CMB observations in the standard cosmology. It is
the inclusion of EDE that lifts the CMB-lockdown on such ULA. Furthermore, we found such ULA can alleviate the tension between AdS-EDE and LSS data, yielding $H_0=72.53_{-0.58}^{+0.51}$ and $S_8=0.785_{-0.016}^{+0.015}$, both within $1\sigma$ range of the locally measured values. There is still residual inconsistency between AdS-EDE+ULA and the EFT likelihood, highlighting the importance of more precise full shape matter power spectrum data, e.g. DES-Y3 \cite{DES:2021wwk}, in studying DM and EDE, which might worth further research in future studies.

It is interesting that $r_a\sim6\%$ always implies
$f_a\sim10^{17}$ GeV (despite the axion mass or thawing time),
which fits well with the string theory expectation
\cite{Svrcek:2006yi,Arvanitaki:2009fg}, and AdS vacua are also
ubiquitous in string landscape \cite{Obied:2018sgi}, see also \cite{Garg:2018reu}. Furthermore, in addition to the $10^{-26}\text{eV}$ ULA, the EDE itself might be realized by $10^{-27}\text{eV}$ ULA by exploiting the weak gravity conjecture \cite{Kaloper:2019lpl}.
Thus how to embed relevant models into a UV-complete
theory will be worth exploring, see e.g \cite{Marsh:2011gr,Marsh:2012nm}. In Ref.\cite{Ye:2021nej}, it has
been found that in the pre-recombination solutions of the Hubble
tension, the shift of primordial scalar spectral index scales as
${\delta n_s}\simeq 0.4{\delta H_0\over H_0}$, which seems to
suggest a scale-invariant Harrison-Zeldovich spectrum ($n_s= 1$)
for $H_0\sim 73$km/s/Mpc. Here, we confirm this result again, see
Tables-\ref{early_par}, which might have
radical implications on our understanding of the primordial
Universe.

\paragraph*{Acknowledgments}
This work is supported by NSFC, Nos.12075246, 11690021. Some
figures in this paper are plotted with the help of GetDist
\cite{Lewis:2019xzd}. The computations are performed
on the TianHe-II supercomputer and the Xmaris cluster. We thank Alessandra Silvestri for
useful comments and discussions.

\appendix
\section{More MCMC results}

\subsection{Resutls without $S_8$ and/or $M_B$ prior}\label{apdx:mcmc_s8mb}

\begin{figure}
	\centering
	\includegraphics[width=\linewidth]{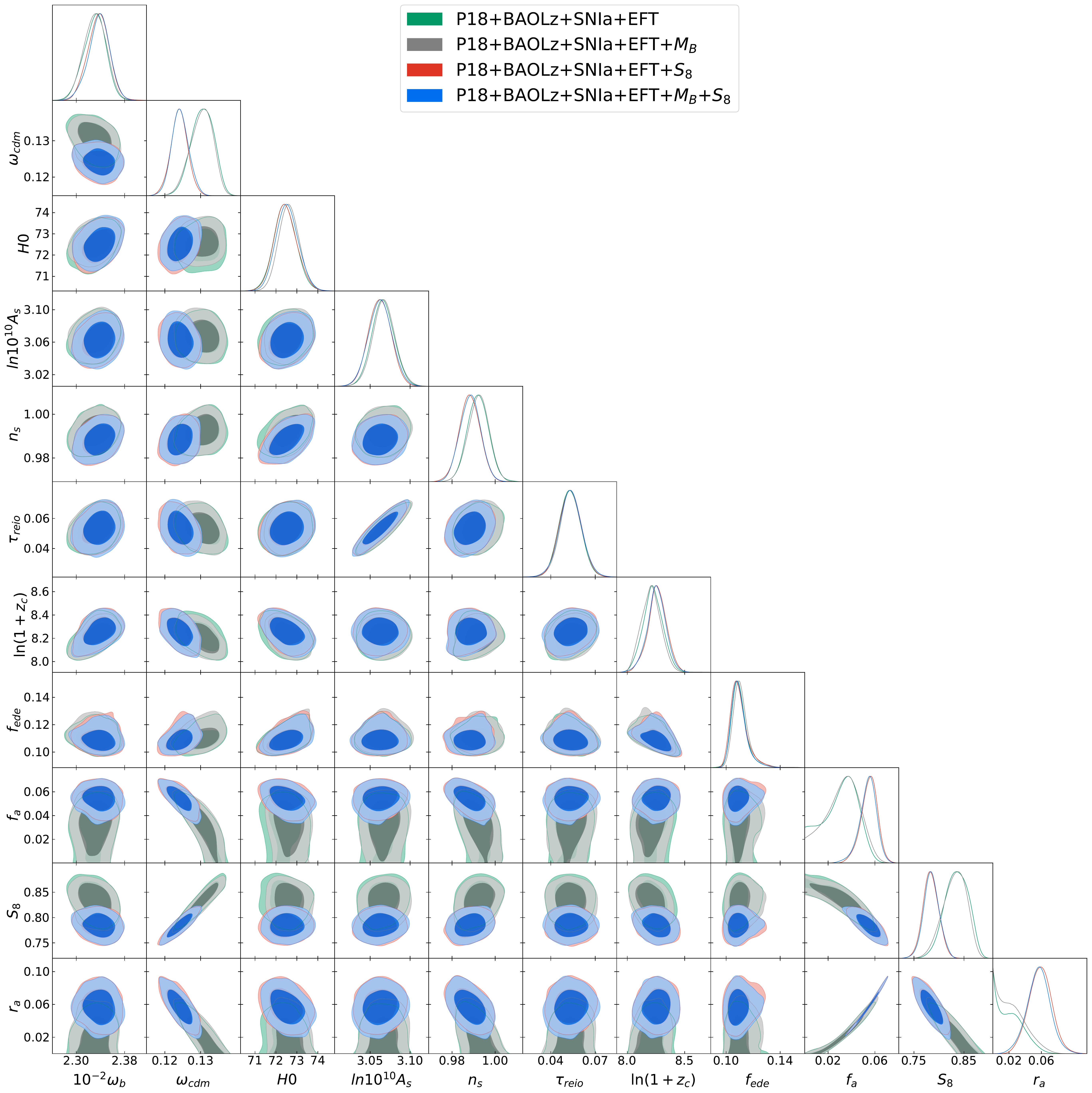}
	\caption{68\% and 95\% posterior distribution of cosmological parameters in AdS-EDE+ULA with and without $S_8$ and/or $M_B$ prior.}
	\label{fig:trig_QDMAP}
\end{figure}

\begin{table}
	\centering
	\begin{tabular}{|c|c|c|c|c|}
		\hline
		&$D_1$&$D_1+M_B$&$D_1+S_8$&$D_1+M_B+S_8$\\ \hline
		$100~\omega{}_{b }$&$2.331(2.329)^{+0.020}_{-0.017}   $& $2.331(2.327)\pm 0.018            $&$2.337(2.342)\pm 0.018            $&$2.338(2.331)_{-0.016}^{+0.018}$\\
		
		$\omega{}_{cdm }$ &$0.1305(0.1312)^{+0.0034}_{-0.0029}$&$0.1303(0.1312)^{+0.0036}_{-0.0027}$& $0.1241(0.1231)\pm 0.0024          $&$0.1242(0.1229)_{-0.0025}^{+0.0023}$\\
		
		$H_0$&$72.44(72.45)\pm 0.53             $ &$72.62(72.44)\pm 0.49             $&$72.46(72.24)^{+0.49}_{-0.57}     $&$72.53(72.14)_{-0.58}^{+0.51}$\\
		
		$\ln10^{10}A_{s }$ &$3.066(3.069)\pm 0.014            $ &$3.067(3.069)\pm 0.014            $&$3.062(3.059)\pm 0.014            $&$3.063(3.066)_{-0.015}^{+0.014}$\\
		
		$n_{s }$& $0.9921(0.9958)\pm 0.0049          $ &$0.9923(0.9959)\pm 0.0048          $&$0.9882(0.989)\pm 0.0048          $&$0.9885(0.9856)_{-0.0048}^{+0.0048}$\\
		
		$\tau{}_{reio }$ & $0.0529(0.057)\pm 0.0072          $ &$0.0536(0.057)\pm 0.0073          $&$0.0532(0.0501)\pm 0.0073          $&$0.0534(0.0549)_{-0.0076}^{+0.007}$\\
		
		\hline
		$\ln(1+z_{c})$&$8.225(8.213)\pm 0.089            $&$8.207(8.213)\pm 0.084            $&$8.261(8.307)^{+0.082}_{-0.11}    $&$8.255(8.256)_{-0.069}^{+0.088}$\\
		
		$f_{ede }$&$0.1098(1.054)^{+0.0038}_{-0.0060}$&$0.1111(0.1053)^{+0.0034}_{-0.0067}$&$0.1095(0.101)^{+0.0037}_{-0.0071}$&$0.1094(0.1124)_{-0.0064}^{+0.0036}$\\
		
		\hline
		$f_{a }/M_p$ &$0.031(0.024)^{+0.019}_{-0.011}   $&$0.032(0.024)^{+0.017}_{-0.012}   $&$0.0550(0.0596)^{+0.0076}_{-0.0059}$&$0.0541(0.0631)_{-0.0057}^{+0.0081}$\\
		
		$r_a$&$< 0.029(0.011)                  $&$< 0.031(0.011)                 $&$0.058(0.0668)\pm 0.014            $&$0.0567(0.0751)_{-0.013}^{+0.015}$\\
		
		\hline
		$M_{B }$&$-19.274(-19.275)\pm 0.015          $ &$-19.269(-19.275)\pm 0.014          $&$-19.274(-19.28)^{+0.013}_{-0.015} $ &$-19.27(-19.28)_{-0.016}^{+0.014}$\\
		
		$S_8$& $0.835(0.847)^{+0.026}_{-0.022}   $ &$0.832(0.847)^{+0.027}_{-0.020}   $&$0.784(0.777)\pm 0.016            $&$0.785(0.772)_{-0.016}^{+0.015}$\\
		
		$\Omega_{m }$ & $0.2998(0.2976)\pm 0.0055          $&$0.2983(0.2975)\pm 0.0052          $&$0.2983(0.3009)\pm 0.0054          $&$0.2975(0.3037)_{-0.0057}^{+0.0053}$\\
		
		\hline
	\end{tabular}
	\caption{68\%C.L. posterior constraints or 95\%C.L. upper bounds on cosmological parameters in AdS-EDE+ULA. $D_1$ denotes the dataset Planck18+BAOLz+SNIa+EFT.}
	\label{Tab:Q_DMAP}
\end{table}

\begin{table}
	\begin{tabular}{|c|c|c|c|c|}
		\hline
		&$D_1$&$D_1+M_B$&$D_1+S_8$&$D_1+M_B+S_8$\\ \hline
		Planck18 high-$l$ TTTEEE&2350.43&2350.54&2353.9&2352.72\\
		\hline
		Planck18 low-$l$ TT&20.11&20.14&20.55&20.71\\
		\hline
		Planck18 low-$l$ EE&396.25&396.26&395.78&396.06 \\
		\hline
		Planck18 lensing&9.95&9.98&9.79&9.82 \\
		\hline
		Pantheon&1026.69&1026.69&1026.88&1026.94 \\
		\hline
		BAO low $z$&2.09&2.10&1.75&1.53 \\
		\hline
		eft withbao highzNGC&66.98&66.95&68.85&69.72\\
		\hline
		eft withbao highzSGC&63.23&63.27&60.89&60.73\\
		\hline
		eft withbao lowzNGC&70.64&70.67&71.24&71.26\\
		\hline
		$S_8$&-&-&1.22&1.18 \\
		\hline
		$M_B$&-&0.65&-&0.79 \\
		\hline
		\hline
		$\chi^2_{CMB}$&2776.74&2776.92&2780.02&2779.31\\
		$\chi^2_{EFT}$&200.85&200.89&200.98&201.71\\
		$\chi^2_{tot}$&4006.36&4007.26&4010.85&4011.46\\
		\hline
	\end{tabular}
	\caption{Bestfit $\chi^2$ per dataset for AdS-EDE+ULA. $D_1$ denotes Planck18+BAOLz+SNIa+EFT.}
	\label{Tab:Q_DMAP chi2}
\end{table}

Denoting Planck18+BAOLz+SNIa+EFT as $D_1$, we add the $S_8$, $M_B$ prior or both as $D_2$ to see their effects on parameter constraints in the AdS-EDE+ULA model with $m_a\simeq1.8h\times10^{-26}$eV and $\alpha_{ads}=3.79\times10^{-4}$ and compute the $Q_{DMAP}$ values. Posterior distributions and bestfit values are shown in Fig.\ref{fig:trig_QDMAP} and Table.\ref{Tab:Q_DMAP} while Table.\ref{Tab:Q_DMAP chi2} reports the corresponding bestfit $\chi^2$ per dataset. The results imply that the $S_8$ prior is crucial in the detection of ULA. This is to be expected, because though EDE opens up the parameter space for ULA, it does not require the existence of ULA since EDE models alone are known to be compatible with CMB+SNIa+BAO. It is the inclusion of $S_8$ information that sets the "need" for ULA due to the exacerbated $S_8$ tension in original EDE models.

\subsection{Results of varying model parameters $m_a$, $\Theta_i$ and $\alpha_{ads}$}\label{apdx:mcmc_releasepar}

\begin{figure}
	\centering
	\includegraphics[width=\linewidth]{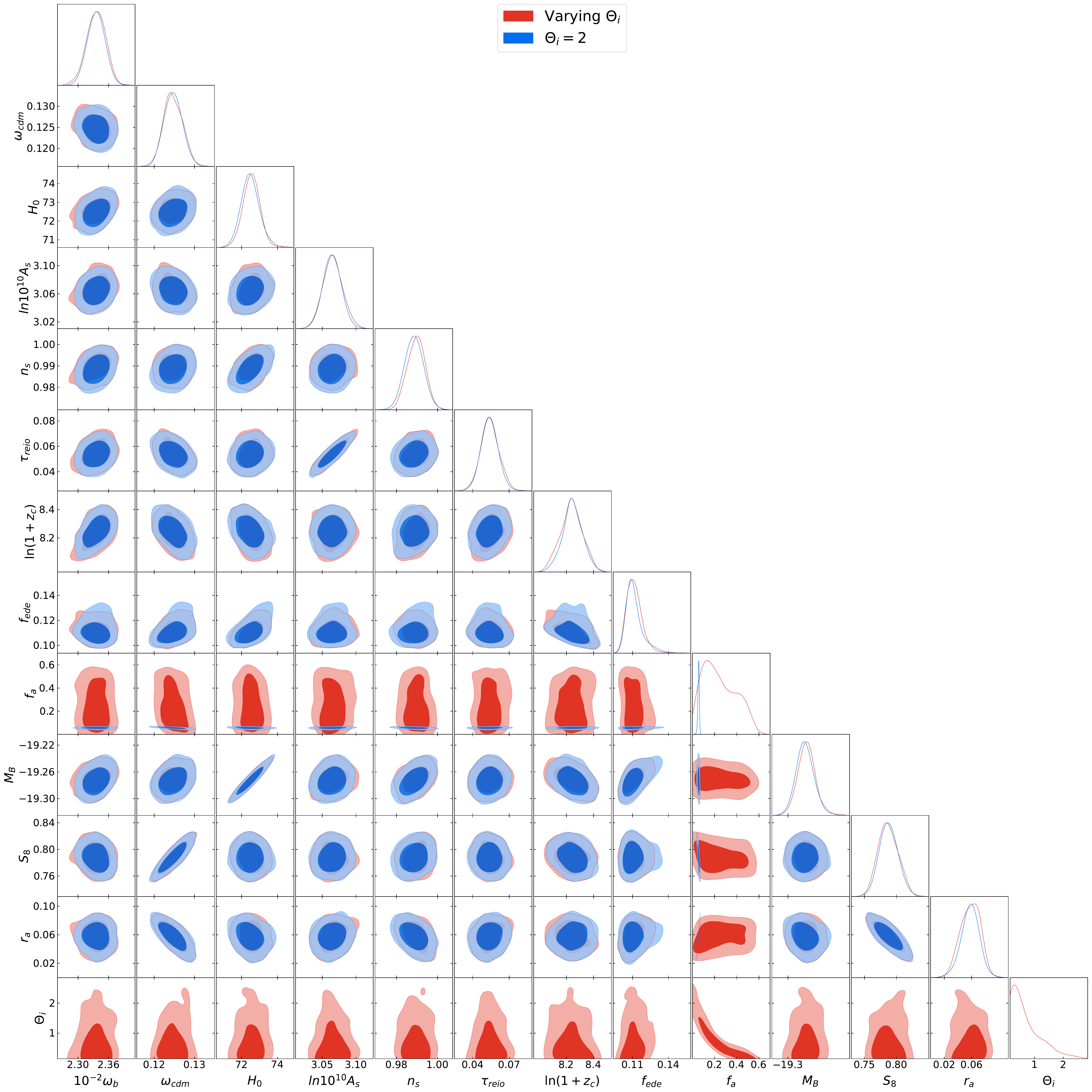}
	\caption{68\% and 95\% posterior distribution of cosmological parameters in AdS-EDE+ULA with initial ULA configuration $\Theta_i$ varied.}
	\label{fig:trig_theta}
\end{figure}

\begin{figure}
	\centering
	\includegraphics[width=\linewidth]{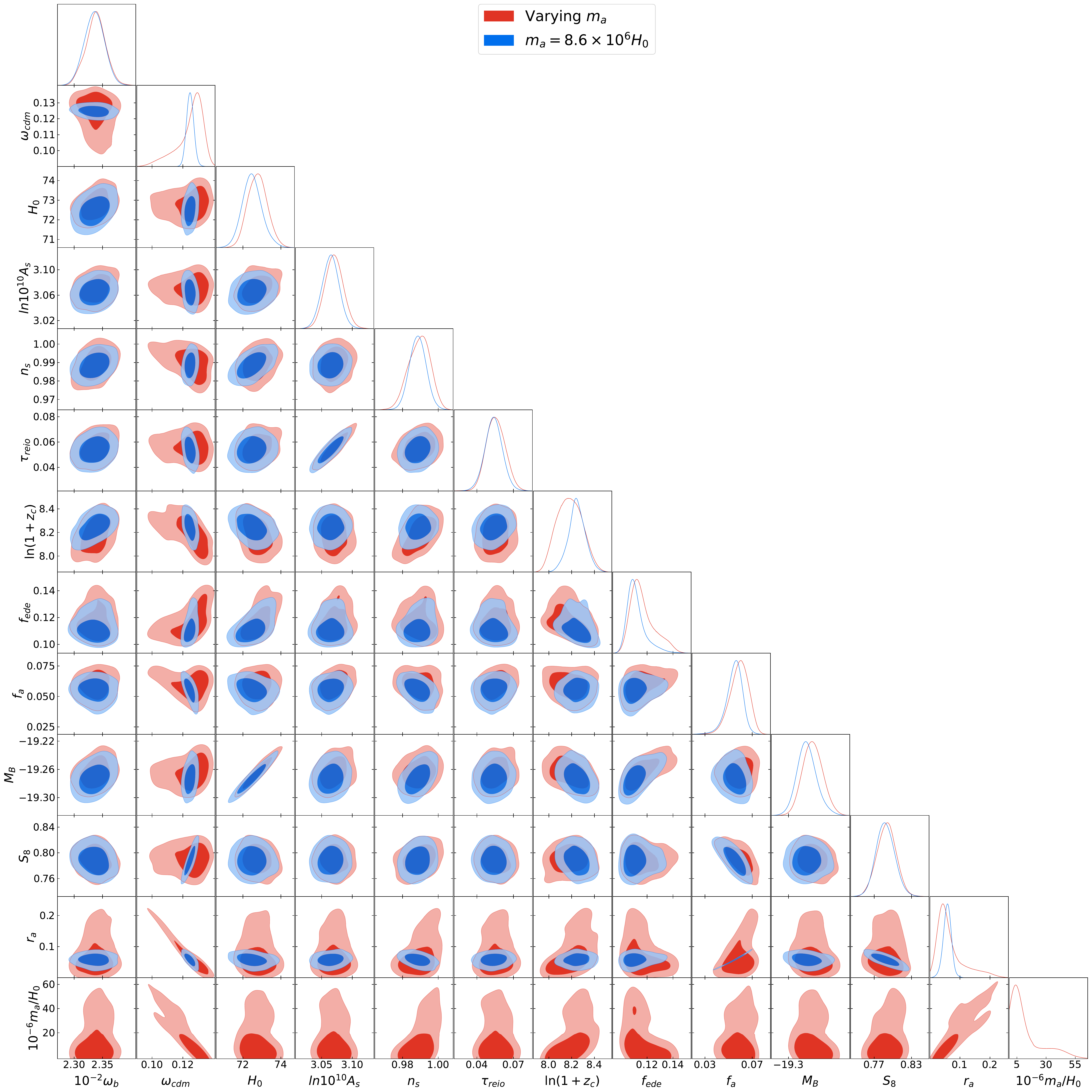}
	\caption{68\% and 95\% posterior distribution of cosmological parameters in AdS-EDE+ULA with ULA mass $m_a$ varied.}
	\label{fig:trig_ma}
\end{figure}

\begin{figure}
	\centering
	\includegraphics[width=\linewidth]{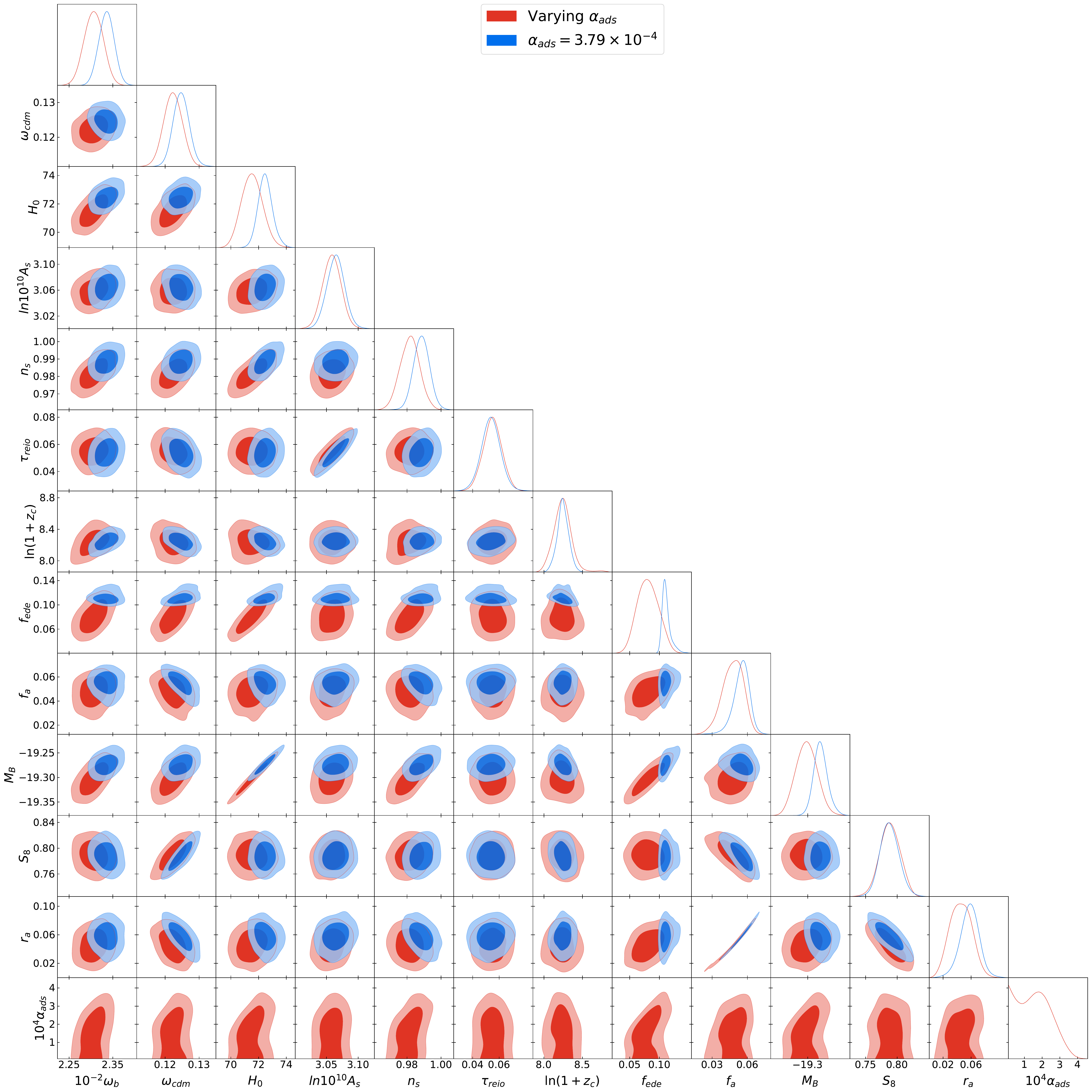}
	\caption{68\% and 95\% posterior distribution of cosmological parameters in AdS-EDE+ULA with AdS relative depth $\alpha_{ads}$ varied.}
	\label{fig:trig_ads}
\end{figure}

\begin{table}
	\begin{tabular}{|c|c|}
		\hline
		Parameter&Prior\\
		\hline
		$10^{-6}m_a/H_0$&$[10^{-1},10^2]$\\
		\hline
		$\Theta_i$&$[0.15,3]$\\
		\hline
		$10^4\alpha_{ads}$&$>0.1$\\
		\hline
	\end{tabular}
	\caption{Flat priors on $\{m_a, \Theta_i, \alpha_{ads}\}$.}
	\label{Tab:prior}
\end{table}
In the main text we fixed three model parameters $\{m_a,\Theta_i,\alpha_{ads}\}$ of AdS-EDE+ULA due to convergence or volume effect reasons. To illustrate their effects on cosmological parameters, in this subsection we present MCMC results of AdS-EDE+ULA with $m_a$, $\Theta_i$ and $\alpha_{ads}$ letting free to vary one by one. Due to the considerably prolonged convergence time, we use the dataset Planck18+fsBAO+SNIa+$S_8$+$M_B$ excluding the slow EFT data. As noted in the main text and Tab.\ref{early_par}, this dataset produces very similar cosmological constraints as that with EFT but runs much faster. We use flat priors on $\{m_a, \Theta_i, \alpha_{ads}\}$, see Tab.\ref{Tab:prior}. We constrain axion mass $m_a$ to roughly the mass range $10^{-24}\sim10^{-28}$eV which is wide enough to cover the interesting mass range $\mathcal{O}(10^{-26}\rm{eV})\gtrsim m_a\gtrsim\mathcal{O}(10^{-27}\rm{eV})$ as argued in section-\ref{sec:result} and also excludes regions where the scalar field behaves like (early) dark energy or is degenerate with cold dark matter. For initial phase $\Theta_i$ we exclude the extreme values near $0$ or $\pi$. For $\alpha_{ads}$ we set a small lower bound so that we are sampling models with an AdS phase but variable depth. Note that models with $\alpha_{ads}$ free to take both positive and negative values have been studied in Ref.\cite{Ye:2020oix} and it is concluded the existence of the AdS phase is weakly hinted. Posterior distributions of cosmological parameters after releasing $\{m_a, \Theta_i, \alpha_{ads}\}$ separately are plotted in Fig.\ref{fig:trig_theta}, Fig.\ref{fig:trig_ma} and Fig.\ref{fig:trig_ads}. For all plots we also include the AdS-EDE+ULA with $\{m_a, \Theta_i, \alpha_{ads}\}$ fixed to their values in the main text (i.e. the last column of Tab.\ref{early_par}) as reference.

Fig.\ref{fig:trig_theta} plots the results of varying $\Theta_i$. ULA fraction $r_a$ is determined by its initial energy fraction at thawing to a good extent because after that ULA energy density redshifts just as matter. Since data constrained $r_a$ to a specific range, it is expected that $f_a$ and $\Theta_i$ are strongly anti-correlated when $m_a$ fixed because the energy fraction at thawing is proportional to $V(\Theta_i)= m_a^2f_a^2(1-\cos\Theta_i)$. According to Fig.\ref{fig:trig_theta}, the data constraint on $r_a$ is robust against varying $\Theta_i$. $r_a$ only constrains the value of a specific combination of $f_a$ and $\Theta_i$ when $m_a$ is fixed, thus $\Theta_i$ and $f_a$ alone are poorly constrained because of the strong anti-correlation. Fixing $\Theta_i$ amounts to picking out a spefic value of $f_a$ and does not bias our conclusion about $r_a$.

Fig.\ref{fig:trig_ma} plots the results of varying $m_a$. Similar to the case of $\Theta_i$, varying $m_a$ relaxes the constraints on $f_a$ because $V(\Theta_i)\propto m_a^2f_a^2$. What is different is that constraints on $r_a$ is no longer stable against varying $m_a$. This is to be expected given the arguments in the main text that EDE only opens the parameter space for ULA in the mass range $\mathcal{O}(10^{-26}\rm{eV})\gtrsim m_a\gtrsim\mathcal{O}(10^{-27}\rm{eV})$ and the results of Ref.\cite{Hlozek:2017zzf,Lague:2021frh}. Moreover, the heavier ULA is, the more degenerate it is with CDM for the datasets we considered, thus the $m_a$ distribution shows a long tail in the high mass range. In turn, when ULA is indistinguishable from CDM, large portion of CDM can be replaced by ULA which is also reflected in Fig.\ref{fig:trig_ma} as relaxed bounds on $r_a$ and $\omega_{cdm}$ and anti-correlation between them.

\begin{figure}
	\centering
	\includegraphics[width=0.8\linewidth]{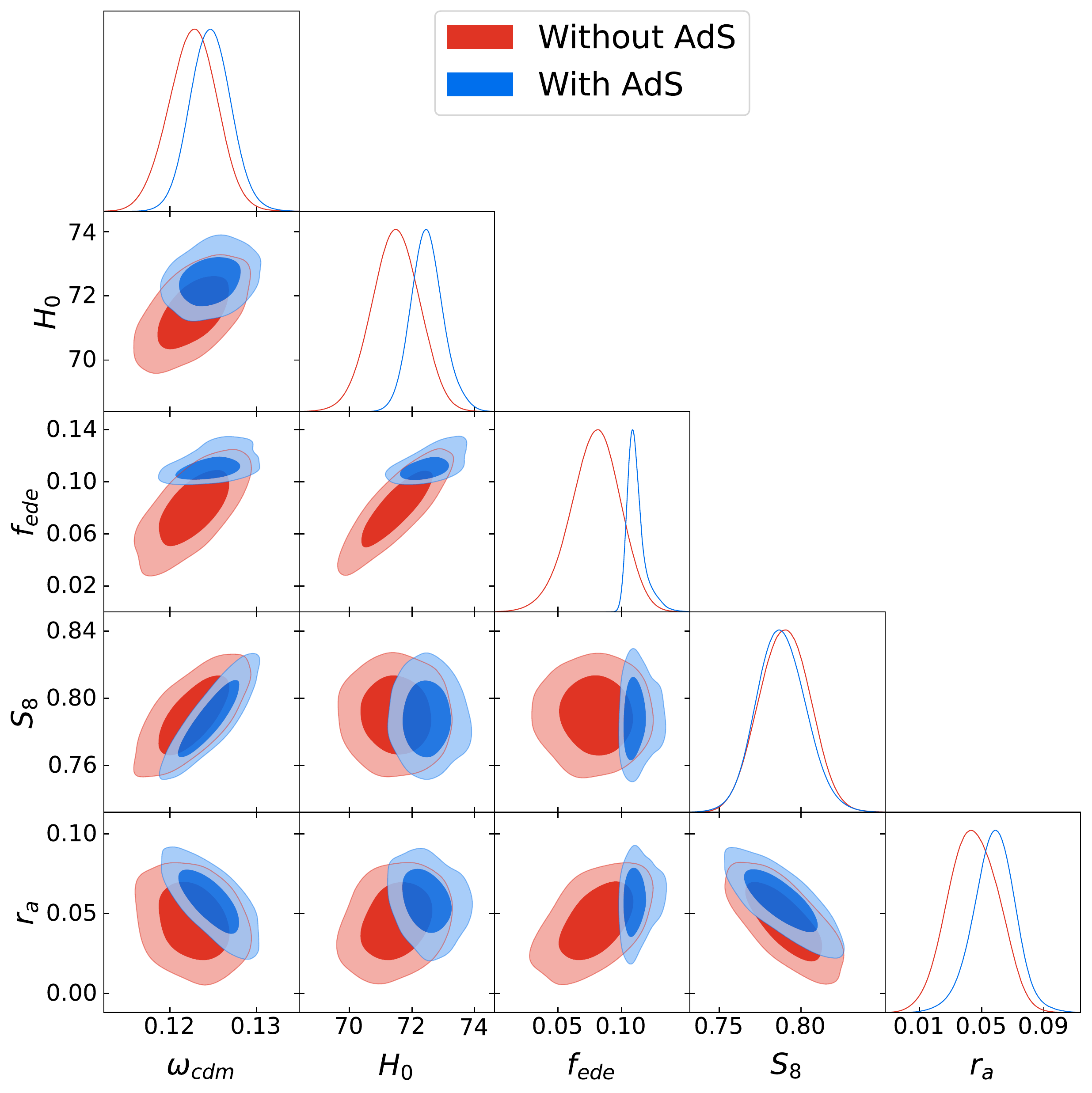}
	\caption{68\% and 95\% posterior distribution of cosmological parameters in AdS-EDE+ULA with and without an AdS phase.}
	\label{fig:trig_adsra}
\end{figure}

\begin{table}
	\begin{tabular}{|c|c|c|}
		\hline
		Dataset&$\alpha_{ads}=0$&$\alpha_{ads}=3.79\times10^{-4}$\\
		\hline
		Planck18&2777.43&2781.03 \\
		\hline
		Pantheon&1026.48&1026.89 \\
		\hline
		fsBAO&8.03&9.25 \\
		\hline
		$S_8$&4.27&1.06 \\
		\hline
		$M_B$&2.89&0.12 \\
		\hline
		\hline
		$\chi^2_{tot}$&3819.09&3818.35\\
		\hline
	\end{tabular}
	\caption{Bestfit $\chi^2$ for AdS-EDE+ULA with $\alpha_{ads}=0$ and $\alpha_{ads}=3.79\times10^{-4}$.}
	\label{tab:alphachi2}
\end{table}

Fig.\ref{fig:trig_ads} plots the results of varying $\alpha_{ads}$.  Fig.\ref{fig:trig_adsra} and Table.\ref{tab:alphachi2} further compares AdS-EDE+ULA with $\alpha_{ads}=0$ (no AdS phase, rolling EDE potential) and $\alpha_{ads}=3.79\times10^{-4}$ in terms of constraints on relevant cosmological parameters and bestfit $\chi^2$. Despite fitting the local Hubble measurement better without worsening the overall fit, models with non-negligible AdS phase is punished by the phase space volume effect. Non-zero $\alpha_{ads}$ value puts a theoretical lower bound on $f_{ede}$ and cuts out part of the viable parameter space due to the fact that the field would be trapped in the disastrous negative energy region if $f_{ede}$ becomes too small. Thus the smaller $\alpha_{ads}$ is, the larger the viable phase space is. As a result, MCMC chain will tend to explore points near $\alpha_{ads}=0$ despite better fit at $\alpha_{ads}\ne0$, making AdS only weakly hinted. Fig.\ref{fig:trig_ads} is also compatible with the results of Ref.\cite{Ye:2020oix}. Constraints on $r_a$ is less strict when $\alpha_{ads}$ is let free to vary or fixed to $0$ but the 95\%C.L. detection conclusion is unchanged. As argued in the main text, it is EDE that opens up the parameter space for ULA. Thus the higher $f_{ede}$ the more ULA is allowed. Because the existence of an AdS phase increases $f_{ede}$, it also increases the allowed ULA fraction $r_a$. However, since $\alpha_{ads}$ is not directly related to the mechanism that opens up ULA parameter space, it is thus not required by the detection of ULA at 95\%C.L.

\bibliography{ref}

\end{document}